# Algorithmic Addiction by Design: Big Tech's Leverage of Dark Patterns to Maintain Market Dominance and its Challenge for Content Moderation

**Michelle Nie**[1]
[1]Independent Researcher
michellesnie@gmail.com

**Abstract**

Today's largest technology corporations, especially ones with consumer-facing products such as social media platforms, use a variety of unethical and often outright illegal tactics to maintain their dominance. One tactic that has risen to the level of the public consciousness is the concept of addictive design, evidenced by the fact that excessive social media use has become a salient problem, particularly in the mental and social development of adolescents and young adults. As tech companies have developed more and more sophisticated artificial intelligence (AI) models to power their algorithmic recommender systems, they will become more successful at their goal of ensuring addiction to their platforms. This paper explores how online platforms intentionally cultivate addictive user behaviors and the broad societal implications, including on the health and well-being of children and adolescents. It presents the usage of addictive design — including the usage of dark patterns, persuasive design elements, and recommender algorithms — as a tool leveraged by technology corporations to maintain their dominance. Lastly, it describes the challenge of content moderation to address the problem and gives an overview of solutions at the policy level to counteract addictive design.

## Big Tech and Market Power

The largest technology corporations today — Google, Amazon, Meta, Microsoft, and Apple — which can collectively be described as "Big Tech" (Birch & Bronson, 2022), have been embroiled in an all-out war to secure market power, beat each other in a number of industries, and edge out would-be challengers. In order to understand the incentives of Big Tech, one needs to take an ecosystem-level perspective.

Big Tech firms tightly control vast ecosystems of digital services and devices, including social media platforms, search engines, smartphones, operating systems, browsers, cloud computing, and generative AI models (Von Thun & Hanley, 2024).

Over the years, Big Tech has built mega-popular consumer-facing applications and products that have scaled rapidly and captured markets through "platformization" and network effects (Birch & Bronson, 2022). Through these applications and products, which reach billions of users around the world, Big Tech has amassed troves of valuable, proprietary user data. It harvests profits through ads. And it reaches consumers that drive their revenue sources through commercial gateways, one of the largest being social media. Indeed, a majority of the top social media and online communications applications — including Facebook, Instagram, YouTube, LinkedIn, WhatsApp, and Messenger — are linked to a Big Tech corporation.

Social media platforms are a major way companies push ads to users and promote other products through self-leveraging, such as generative AI models — for instance, Meta promotes usage of and nurtures dependency on its AI chatbot by integrating it within Facebook and Instagram (Von Thun & Hanley, 2024).

## Overview of Addictive Design Elements and their Societal Impact

The addictive design elements described in this section be described using the term "dark patterns" — defined by Luguri and Strahilevitz (2021) as "user interfaces whose designers knowingly confuse users, make it difficult for users to express their actual preferences, or manipulate users into taking certain actions."

### Personalized content algorithms or "recommender systems"

One main strategy to capture user attention and encourage addiction are interaction-based recommender systems, in particular personalized systems that are designed to keep users on the platform as long as possible, consume more advertisements, and generate maximum profits for tech companies. These systems are used on Instagram's Explore page and TikTok's For You Page. Studies using brain scans have shown that personalized content, such as TikTok videos, trigger stronger activity in areas of the brain associated with addiction as compared to non-personalized videos (Su et al., 2021). In addition, it has even been shown that recom-mender systems can exploit users' data to personalize content with their fears and vulnerabilities, harming their health and wellbeing (Panoptykon Foundation, 2023).

### Infinite scroll and autoplay

These features create a continuous flow of content, getting rid of a natural stopping point for users. The automatic loading of new content keeps users engaged and delivers a constant flow of dopamine to the brain, removing the need to actively decide what to browse next and removing their agency and autonomy while using social media platforms. Examples include YouTube and Netflix's autoplay features, Instagram's infinite scroll, and Spotify's Radio function.

### Exploiting psychosocial needs

Design features intentionally play into users' psychological needs, vulnerabilities, and desires such as social belonging, social anxiety, and fear of missing out (FOMO). Features like ephemeral Instagram Stories encourage constant engagement and increase pressure to be permanently online, increasing the risk of stress and online burnout (Committee on the Internal Market and Consumer Protection, 2023).

### The age of hyper-personalization powered by AI

Machine learning algorithms have long been used in recommender systems to "learn" users' preferences and personalize social media feeds according to these preferences. (Gillespie, 2022; Hao, 2021). The algorithmic recommender systems were widely known to be one of the most "primitive" applications of artificial intelligence (AI) used in consumer applications. However, over the last few decades, Big Tech has exponentially increased both the sophistication of their AI algorithms and the amount of proprietary data they collect on their users (Maslej, 2025; Radsch, 2024; Birch et al., 2021), which leads to exponential advances in the accuracy of recommender systems and their ability to hyper-personalize social media products (Zhao et al., 2024).

## Algorithmic Addiction and Well-Being

The above behaviors have led to algorithmic addiction, which affects the well-being of all users, but particularly children, whose brain development processes are negatively impacted by extensive use of applications using addictive design such as social media platforms.

In children, problematic smartphone use driven by addictive design is linked to lower life satisfaction and increased mental health symptoms, such as anxiety and depression. This addiction also leads to a decrease in sleep quality and quantity and increased sedentary time, which are both linked to a host of negative physical and mental health consequences (Al-Samarraie et al., 2022). The addictive design of online services leads to increased pressure to perform and social pressure to be permanently online and connected, elevating the risk of stress and burnout.

Problematic use of social media platforms has also been linked to lowered social abilities, such as the ability to connect with family and friends and forge real-world social relationships (Haidt, 2024). Online networks are not as binding or satisfying as stable real-world communities, leading to lower social satisfaction.

## The Challenge for Content Moderation and Policy Solutions

Companies are well aware of the dangers posed by their intentionally addictive social media platforms, as exemplified by Facebook whistleblowers Frances Haugen and Sarah Wynn-Williams (Hao, 2021; U.S. Senate Committee on the Judiciary, 2025). They are also aware that safer recommender systems are possible, such as those based on chronological order or those with more user control over the content displayed, but these alternatives are less profitable for social media platforms.

The challenge to content moderation is twofold. First, content moderators are beholden to the moderation policies set at the corporate or platform level (Singhal et al., 2023), and companies that own these platforms are unwilling to change any policies that would erode their profits or undermine their carefully constructed ecosystems. Second, the overall trend of content moderation, which peaked during the COVID-19 era due to the need to curb the spread of misinformation and disinformation, is in decline. The trend stems from X's (formerly Twitter's) decision in 2022 to loosen content restrictions and has culminated in Mark Zuckerberg's infamous announcement in January that Meta platforms will remove speech restrictions and end fact-checking programs.

Meanwhile, Big Tech faces unprecedented pressure from influential figures in the Trump administration — since they are all headquartered in the United States — to prioritize "free speech" and end "censorship." Examples include JD Vance's lamenting of the "retreat" of free speech at the Munich Security Conference and Federal Trade Commission Chair Andrew Ferguson's launch of an investigation of Big Tech policies that censor Americans. In such a political environment, it is highly unlikely that Big Tech will voluntarily impose stricter content policies on its platforms when they know these drive engagement and are popular with the administration.

However, a survey conducted by University of Oxford and the Technical University of Munich shows that the majority of surveyed users support content moderation on social media platforms (Theocharis et al., 2025). Bans on social media use for children and adolescents are on the rise, including Australia's social media ban for users under 16 and the United States' proposed Kids Off Social Media Act (Australian eSafety Commissioner, 2025; Schatz, n.d.). Although banning children from social media outright will not solve the underlying issues and incentives around addictive design, these initiatives demonstrate that action must be taken urgently to combat algorithmic addiction.

Platforms must be forced to implement less harmful, less addictive designs and strategies through regulation. The push for more regulation has been driven by the European Union, where MEP Kim Van Sparrentak has led the charge for the European Commission to introduce new legislation specifically targeted towards addictive design. A report chaired by Van Sparrentak states that the current digital regulatory framework in the European Union is insufficient to combat additive design (Committee on the Internal Market and Consumer Protection, 2023). It states that, for example, the Digital Markets Act (DMA) and Digital Services Act (DSA) introduce provisions against the use of dark patterns, but neither addresses addictive design directly. Adopted by the European Parliament, the report highlighted the urgent need for specific legislation targeting addictive design.

Anti-addictive design legislation could include provisions requiring social media platforms to turn off attention-seeking features by default, to implement pagination instead of scrolling by default, demoting harmful or addictive content, and promoting and prioritizing alternative recommender systems based on chronological order or increased user control (Gillespie, 2022; Campbell et al., 2013). These remedies, if implemented, would offer a healthier "content diet" that is less likely to be addictive.

In the United States, Section 230 of the Communications Decency Act (known simply as Section 230) must be amended (Protection for Private Blocking and Screening of Offensive Material, 2011). Section 230 states that no provider or user of an interactive computer service shall be held liable as the publisher or speaker of content created by another information content provider. This piece of the United States Code has single-handedly enabled all the harms stemming from social media platforms — including addiction, abuse, violence — since social media companies could not be held liable to the responsibilities shouldered by publishers (Citron, 2023).

Furthermore, applying an ecosystem perspective to the problem also allows us to see that the issue of addictive design is directly connected with Big Tech's all-consuming drive to achieve market dominance. Tactics such as vertical integration via mergers and acquisitions, self-preferencing, and bundling and tying have all allowed Big Tech to create an ecosystem and a virtuous cycle that allows it to become more and more dominant. By applying competition policy — including antitrust regulation, merger control procedures, and ex-ante digital regulation (such as the DMA in the EU and the Digital Markets, Competition and Consumers Act in the UK) — competition enforcers can apply remedies up to and including the structural separation of social media platforms from the rest of the companies' business, chipping away at the incentives that drive Big Tech's push towards hyper-addictive design.

## Conclusion

This position paper has explored how Big Tech corporations deliberately design addictive features into their platforms to maintain market dominance and maximize profits. Through dark patterns, persuasive design elements, and increasingly sophisticated AI-powered recommender systems, these companies have created digital ecosystems that prioritize user engagement over well-being. The societal implications of these practices are particularly severe for children and adolescents, who face increased risks of physical and mental health issues, as well as decreased social capacities.

While content moderation alone cannot address these structural problems, a multi-faceted approach combining regulation, antitrust enforcement, and technical solutions offers promising paths forward. Ultimately, enforcement of antitrust law may be necessary to realign incentives and break the cycle of harmful engagement-maximizing design.

### Limitations and counterarguments

A significant limitation of the approach outlined in this paper is the challenge of implementation of regulation or structural remedies in an increasingly polarized political environment. Deregulation of the digital economy is on the rise, led by the U.S. administration and followed by the UK and perhaps the EU (Foy & Moens, 2025). Opponents of regulation argue that it could stifle innovation or constitute censorship.

Additionally, while this paper has focused on Big Tech's responsibility, it gives limited attention to user agency. Some argue that education and awareness campaigns could empower users to make more informed choices about their technology use — the U.S. Surgeon General's Advisory on Social Media and Youth Mental Health (2023) even includes actions that parents, caregivers, and even children can take to set boundaries with platforms and educate themselves about the harms of social media. However, this perspective underestimates the ecosystem-level nature of the problem, the malicious psychological techniques employed by social media platforms, and the overwhelming power imbalance between individual users and multi-billion dollar corporations.

Despite these challenges, the mounting evidence of harm caused by algorithmic addiction demands action. As companies continue developing even more sophisticated AI-powered recommendation systems, and as they come under increasing pressure from shareholders to realize profits, the potential for exploitation and harm only increases. By addressing the root causes — the market incentives and business models that profit from addictive design — rather than merely tackling the surface-level problems, we can work toward digital ecosystems that enhance rather than undermine well-being and autonomy.